\documentclass{llncs}

\usepackage[bookmarks = true, colorlinks=true, linkcolor = black, citecolor = black, menucolor = black, urlcolor = black]{hyperref}
\usepackage{graphicx}
\usepackage{amssymb}
\usepackage{color}
\usepackage{listings}
\usepackage{tikz}
\usepackage{cite}

\begin{document}

\title{Verifying a plaftorm for digital imaging: \\ a multi-tool strategy\thanks{Partially supported by Ministerio de Educaci\'on y Ciencia, project MTM2009-13842-C02-01, and by the European Union's
7th Framework Programme under grant agreement nr. 243847 (ForMath). The final publication is available at \url{http://link.springer.com}.}}

\titlerunning{Verifying a plaftorm for digital imaging}  
%
\author{J\'onathan Heras\inst{1}  \and Gadea Mata\inst{2} \and Ana Romero\inst{2} \and Julio Rubio\inst{2}  \and  \\ Rub\'en S\'aenz\inst{2}}
%
%
%

\institute{ School of Computing, University of Dundee, UK \and Department of Mathematics and Computer Science, University of La Rioja, Spain \\
\email{jonathanheras@computing.dundee.ac.uk, gadea.mata@unirioja.es, ana.romero@unirioja.es, julio.rubio@unirioja.es,ruben.saenz@unirioja.es}}

\maketitle              

\begin{abstract}

Fiji is a Java platform widely used by biologists and other
experimental scientists to process digital images. In our research, 
made together with a biologists team, we use Fiji
in some pre-processing steps before undertaking a homological
digital processing of images. In a previous work, we have formalised
the correctness of the programs which use homological techniques to
analyse digital images. However, the verification of Fiji's pre-processing
step was missed. In this paper, we present a \emph{multi-tool} approach
(based on the combination of Why/Krakatoa, Coq and ACL2) filling this gap.

\end{abstract}


\section{Introduction}

Fiji~\cite{Fiji} is a Java platform widely used by biologists and other
experimental scientists to process digital images. In our research, made together with a biologists team,
we use Fiji in some pre-processing steps before undertaking a homological
digital processing of images.

Due to the fact that the reliability of results is instrumental in biomedical research, 
we are working towards the certification of the programs
that we use to analyse biomedical images -- here, certification means
verification assisted by computers. In a previous work, see~\cite{Calculemus2012,TCL},
we have formalised two homological techniques to process biomedical images. However, in both cases,
the verification of Fiji's pre-processing step was not undertaken.

%

Being a software built by means of plug-ins developed by several authors, Fiji is messy,
very flexible (program pieces are used in some occasions with a completely different  objective
from the one they were designed), contains many redundancies and
dead code, and so on. In summary, it is a big software system which has
not been devised to be formally verified. So, this endeavour is challenging.

There are several approaches to verify Java code; for instance, proving the correctness of the associated Java bytecode, see~\cite{jvm}.
In this paper, we use Krakatoa~\cite{FM07} to specify and prove the correctness
of Fiji/Java programs. This experience allows us to evaluate both the verification of
\emph{production} Fiji/Java code, and the  Krakatoa tool itself in an unprepared
scenario.

Krakatoa uses some automated theorem provers (as Alt-Ergo~\cite{altergo} or CVC3~\cite{cvc3})
to discharge the proof obligations generated by means of the Why tool~\cite{FM07}. When a proof
obligation cannot be solved by means of the automated provers, the corresponding statement is
generated in Coq~\cite{Coq}. Then, the user can try to prove the missing property by interacting with 
this proof assistant.

In this picture, we add the ACL2 theorem prover~\cite{ACL2}. ACL2 is an automated theorem prover but
more powerful than others. In many aspects, working with ACL2 is more similar to interactive provers than
to automated ones, see~\cite{ACL2}. Instead of integrating ACL2 in the architecture of Why/Krakatoa,
we have followed another path leaving untouched the Why/Krakatoa code. Our approach reuses a proposal presented in~\cite{XLL} to
translate first-order Isabelle/HOL theories to ACL2 through an XML specification
language called XLL~\cite{XLL}. We have enhanced our previous tools to translate Coq theories to the XLL language, and then apply the 
tools developed in~\cite{XLL} to obtain ACL2 files. In this way, we can use, unmodified, the Why/Krakatoa framework;
the Coq statements are then translated (if needed) to ACL2, where an automated proof is tried; if it succeeds,
Coq is only an intermediary specification step; otherwise, both ACL2 or Coq can be interactively used to
complete the proof.

The organization of the paper is as follows. The used tools together with our general way of
working are briefly presented in Section~\ref{sec:context}. Section~\ref{sec:fijikrakatoa} deals with a methodology to ``tame'' production Fiji code
in such a way that it is acceptable for Why/Krakatoa -- this method is general enough to be applied to any Java code.
Section~\ref{sec:spec} describes an example of the kind of specification
we faced. The role of ACL2, and the tools to interoperate between Coq and ACL2, are explained in Section~\ref{sec:role}.
The exposition style along the paper tries to be clear (without much emphasis on formal aspects), driven by real examples extracted from our
programming experience in Fiji; in the same vein, Section~\ref{sec:action} contains a complete example illustrating the
actual role of ACL2 in our setting. The paper ends with a conclusions section and the bibliography.

All the programs and examples presented throughout this paper are available at \url{http://www.computing.dundee.ac.uk/staff/jheras/vpdims/}.


\section{Context, tools, method}\label{sec:context}

\subsection{Context}


Fiji~\cite{Fiji} is a Java program which can be described as a distribution 
of ImageJ \cite{ImageJ}. These two programs help with the research in life
sciences and biomedicine since they are used to process and analyse biomedical
images. Fiji and ImageJ are open source projects and their functionality
can be expanded by means of either a macro scripting language or Java plug-ins.
Among the Fiji/ImageJ plug-ins and macros, we can find functionality which allows
us to binarise an image via different threshold algorithms, homogenise images
through filters such as the ``median filter'' or obtain the maximum projection of
a stack of images.

In the frame of the ForMath European project~\cite{formath}, one of the tasks
is devoted to the topological aspects of digital image processing. The objective
of that consists in formalising enough mathematics to verify programs in the area
of biomedical imaging. In collaboration with the biologists team directed by
Miguel Morales, two plug-ins for Fiji have been developed (SynapCountJ~\cite{synapcount}
and NeuronPersistentJ~\cite{persistent}); these programs are devoted to analyse
the effects of some drugs on the neuronal structure. At the end of such analysis,
some homological processing is needed (standard homology groups in SynapCountJ and persistent
homology in NeuronPersistentJ). As explained in the introduction, we have verified these
last steps~\cite{Calculemus2012,TCL}. But all the pre-processing steps, based
on already-built Fiji plug-ins and tools, kept unverified. This is the gap we try
to fill now, by using the facilities presented in the sequel.

\subsection{Tools}\label{subsec:tools}

\subsubsection{Why/Krakatoa: Specifying and verifying Java code.}

The Why/Krakatoa tools~\cite{FM07} are an environment for proving the correctness of Java programs annotated with JML~\cite{JML}
specifications which have been successfully applied in different context, see~\cite{BPZ12}. The environment involves 
three distinct components: the Krakatoa tool, which reads the annotated Java files and produces
a representation of the semantics of the Java program into Why's input language; the Why tool, which computes proof
obligations (POs) for a core imperative language annotated with pre- and post-conditions, and  finally several automated theorem provers which are included in 
the environment and are used to prove the POs. When some PO cannot be solved by means of the automated provers, corresponding statements are automatically generated in 
Coq~\cite{Coq}, so that the user can then try to prove
the missing properties in this interactive theorem prover. The POs generation is based on a Weakest Precondition calculus and the validity of all generated POs 
implies the soundness of the code with respect to the given specification. The Why/Krakatoa tools are available as open source software at~\url{http://krakatoa.lri.fr}.

\subsubsection{Coq and ACL2: Interactive theorem proving.}

Coq~\cite{Coq} is an interactive proof assistant for constructive higher-order logic based on
the Calculus of Inductive Construction. This system provides a formal language to
write mathematical definitions, executable algorithms and theorems together with an
environment for semi-interactive development of machine-checked proofs. Coq has been successfully
used in the formalisation of relevant mathematical results; for instance, the recently proven Feit-Thompson Theorem~\cite{FOOT}.

ACL2~\cite{ACL2} is a programming language, a first order logic and an automated theorem prover.
Thus, the system constitutes an environment in which algorithms can be defined and executed,
and their properties can be formally specified and proved with the assistance of a mechanical
theorem prover. ACL2 has elements of both interactive and automated provers. ACL2 is automatic 
in the sense that once started on a problem, it proceeds without
human assistance. However, non-trivial results are not usually proved in the first attempt, and the
user has to lead the prover to a successful proof providing a set of lemmas, inspired by the failed proof
generated by ACL2. This system has been used for a variety of important formal methods projects of industrial and
commercial interest~\cite{ACL2-industry} and for implementing large proofs in mathematics.

\subsection{Method}

In this section, we present the method that we have applied to verify Fiji code.
This process can be split into the following steps.

\begin{enumerate}

\item Transforming Fiji code into compilable Krakatoa code.

\item Specifying Java programs.

\item Applying the Why tool.

\item If all the proof obligations are discharged automatically by the provers integrated
in Krakatoa, stop; the verification has ended.

\item Otherwise, study the failed attempts, and consider if they are under-specified; if
it is the case, go again to step (2).

\item Otherwise, consider the Coq expressions of the still-non-proven statements and transform
them to ACL2.

\item If all the statements are automatically proved in ACL2, stop; the verification
has ended.

\item Otherwise, by inspecting the failed ACL2 proofs, decide if other specifications
are needed (go to item (2)); if it is not the case, decide if the missing proofs should
be carried out in Coq or ACL2.

\end{enumerate}

\noindent The first step is the most sensitive one, because it is the only point where informal (or, rather,
semi-formal) methods are needed. Thus, some unsafe, and manual, code transformation can be required. To
minimize this drawback, we apply two strategies:

\begin{itemize}

\item First, only well-known transformations are applied; for instance, we eliminate inheritance
by ``flattening'' out the code, but without touching the real behaviour of methods.

\item Second, the equivalence between the original code and the transformed one is systematically
tested.

\end{itemize}

Both points together increase the reliability of our approach; a more detailed description of
the transformations needed in step (1) are explained in Section~\ref{sec:fijikrakatoa}.
Step (2) is quite well-understood, and some remarks about this step are provided in Section~\ref{sec:spec}.
Steps (3)-(6) are mechanized in Krakatoa. The role of ACL2 (steps (6)-(8)) is explained in Section~\ref{sec:role}
and, by means of an example, in Section~\ref{sec:action}.


\section{Transforming Fiji-Java to Krakatoa-Java}\label{sec:fijikrakatoa}

In its current state, the Why/Krakatoa system does not support the complete Java programming language and has some limitations.
In order to make a Fiji Java program compilable by Krakatoa we have to take several steps.
\begin{enumerate}

\item
Delete annotations. Krakatoa JML annotations will be placed between \texttt{\small {$\backslash$*@}} and \texttt{\small @*$\backslash$}. Therefore, we need 
to remove other Java Annotations preceded by~\texttt{\small @}.

\item Move the classes that are referenced in the file that we want to compile into the directory \emph{whyInstallationDir/java\_api/}. For example, the class
\emph{RankFilters} uses the class \emph{java.awt.Rectangle};
therefore, we need to create the folder \emph{awt} inside the \emph{java} directory that already exists, and put the file \emph{Rectangle.java} into it.
Moreover, we can remove the body of the methods because only the headers and the fields of the classes will be taken into consideration.

We must iterate this process over the classes that we add. The files that we add into the \emph{java\_api} directory can contain \texttt{\small import}, \texttt{\small extends} and
\texttt{\small implements} clauses although the file that we want to compile
cannot do it -- Krakatoa does not support these mechanisms. This is a tough process: for instance, to make use of the class \emph{Rectangle}, we need to add fifteen classes.

\item
Reproduce the behaviour of the class that we want to compile.
Considering that we are not able to use \texttt{\small extends} and \texttt{\small implements} clauses, we need to move the code from the upper classes into the one that we want to
compile in order to have the same behaviour. For instance, the class \emph{BinaryProcessor} extends from \emph{ByteProcessor} and inside its constructor it calls
the constructor of \emph{ByteProcessor}; to solve this problem we need to copy the body of the super constructor at the beginning of the constructor of the class \emph{BinaryProcessor}.

If we find the use of interfaces, we can ignore them and remove the \texttt{\small implements} clause because the code will be implemented in the class
that makes use of the interface.

\item
Remove \texttt{\small import} clauses. We need to delete them from the file that we want to compile and change the places where the corresponding classes appear with the
full path codes. If for example we are trying to use the class \emph{Rectangle} as we have explained in Step 2, we need to replace it by \emph{java.awt.Rectangle}.

\item
Owing to package declarations are forbidden, we need to remove them with the purpose of halting ``\emph{unknown identifier packageName}'' errors.

\item
Rebuild native methods. The Java programming language allows the use of \emph{native} methods, which are written in C or C++ and might be specific 
to a hardware and operating system platform. For example, many of the methods in the class \emph{Math} (which perform basic numeric operations such
as the elementary exponential, logarithm, square root, and trigonometric functions) simply call the equivalent method included in a different class
named \emph{StrictMath} for their implementation, and then the code in \emph{StrictMath} of these methods is just a \emph{native} call.
Since native methods are not written in Java, they cannot be specified and verified in Krakatoa. Therefore, if our Fiji program uses some native methods,
it will be necessary to rewrite them with our own code. See in Section~\ref{sec:action} our implementation (and specification) of the native method 
\texttt{sqrt} computing the square root of a number of type double, based on Newton's algorithm.

\item
Add a clause in \emph{if-else} structures in order to remove ``\emph{Uncaught exception: Invalid\_argument(``equal: abstract value'')}''. We can find an
example in the method \texttt{\small filterEdge} of the class \emph{MedianFilter} where we have to replace the last \emph{else{...}} clause by \emph{else if(true){...}}.

\item
 Remove debugging useless references. We have mentioned in a previous step that we can only use certain static methods that we have manually added to the Why
 core code and therefore we can remove some debugging instructions like \texttt{\small System.out.println(...)}. We can find the usage of standard output printing
 statement in the method \texttt{\small write} of the class \emph{IJ}.

\item
 Modify the declaration of some variables to avoid syntax errors. There can be some compilation errors with the definition of some floats and double values that
 match the pattern \emph{$<$number$>$f} or \emph{$<$number$>$d}. We can see an example in the line $180$ of the file \emph{RankFilters.java}; we have to transform the code:
 \texttt{\small float f = 50f;} into \texttt{\small float f = 50}.

\item
 Change the way that Maximum and Minimum float numbers are written. Those two special numbers are located in the file \emph{Float.java} and there are widely
 used to avoid overflow errors, but they generate an error due to the \emph{eP} exponent. To stop having errors with expressions like \texttt{\small 0x1.fffffeP+127d}
 we need to convert it into \texttt{\small 3.4028235e+38f}.

\end{enumerate}


\section{Specifying programs for digital imaging}\label{sec:spec}

As already said in Section~\ref{subsec:tools}, Fiji and ImageJ are open source projects and many different people from many different teams
(some of them not being computer scientists) are involved in the development of the different Fiji Java plug-ins. This implies that the code
of these programs is in general not suitable for its formal verification and a deep previous transformation process, following the steps
explained in Section~\ref{sec:fijikrakatoa}, is necessary before introducing the Java programs into the Why/Krakatoa system. Even after this initial transformation,
 Fiji programs usually remain complex and their specification in Krakatoa is not a direct process. In this section we present some
examples of Fiji methods that we have specified in JML trying to show the difficulties we have faced.

Once that a Fiji Java program has been adapted, following the ideas of Section~\ref{sec:fijikrakatoa}, and is accepted by the Why/Krakatoa application,
the following step in order to certify its correctness consists in specifying its behaviour (that is, its precondition and its postcondition) by writing
annotations in the Java Modelling Language (JML)~\cite{JML} . The precondition of a method must be a proposition introduced by the keyword
\texttt{\small requires} which is supposed to hold in the pre-state, that is, when the method is called. The postcondition is introduced by the keyword
\texttt{\small ensures}, and must be satisfied in the post-state, that is, when the method returns normally. The notation \texttt{\small $\backslash$result}
denotes the returned value. To differentiate the  value of a variable in the pre- and post- states, we can use the keyword \texttt{\small $\backslash$old} for the pre-state.

Let us begin by showing a simple example. The following Fiji method, included in the class \emph{Rectangle}, translates an object by given horizontal
and vertical increments \texttt{\small dx} and \texttt{\small dy}.

\footnotesize
\begin{verbatim}
/*@	ensures x == \old(x) + dx && y == \old(y) + dy;
  @*/
public void translate(final double dx, final double dy) {
    this.x += dx; this.y += dy;
}
\end{verbatim}

\normalsize
The postcondition expresses that the field \texttt{\small x} is modified
by incrementing it by \texttt{\small dx}, and the field \texttt{\small y} is increased by \texttt{\small dy}.
In this case no precondition is given since all values of \texttt{\small dx} and \texttt{\small dy} are valid, and the
keyword \texttt{\small $\backslash$result} does not appear because the returned type is \texttt{\small void}.

Using this JML specification, Why/Krakatoa generates several lemmas (\emph{Proof Obligations})
which express the correctness of the program. In this simple case, the proof obligations are elementary and they can be easily
discharged by the automated theorem provers Alt-Ergo \cite{altergo} and CVC3 \cite{cvc3}, which are included in the
environment. The proofs of these lemmas guarantee the correctness of the Fiji method \texttt{\small translate} with respect to the given specification.

Unfortunately, this is not the general situation because, as already said, Fiji code has not been designed for its formal verification
and can be very complicated; so, in most cases, Krakatoa is not able to prove the validity of a program from the given precondition and postcondition.
In order to formally verify a Fiji method, it is usually necessary to include annotations
in the intermediate points of the program. These annotations, introduced by the
keyword  \texttt{\small assert}, must hold at the corresponding program point.
For loop constructs (while, for, etc), we must give an \emph{inductive
invariant}, introduced by the keyword \texttt{\small loop\_invariant},
which is a proposition which must hold at the loop entry and be preserved
by any iteration of the loop body. One can also indicate a \texttt{\small loop\_variant}, which must be an expression
of type integer,  which remains non-negative and decreases at each loop iteration,
assuring  in this way the termination of the loop. It is also possible to declare
new logical functions, lemmas and predicates, and to define \emph{ghost variables} which allow one to monitor
the program execution.

Let us consider the following Fiji method included in the class \emph{RankFilters}.
It implements Hoare's find algorithm (also known as \emph{quickselect}) for computing the nth
lowest number in part of an unsorted array, generalizing in this way the computation of the median
element. This method appears in the implementation of the``median filter'',
 a process very common in digital imaging which is used in order to achieve greater homogeneity
 in an image and provide continuity, obtaining in this way a good binarization of the image.

\footnotesize
\begin{verbatim}
/*@ requires buf!=null && 1<= bufLength <= buf.length && 0<=n <bufLength;
  @ ensures Permut{Old,Here}(buf,0,bufLength-1)
  @    && (\forall integer k; (0<=k<=n-1 ==> buf[k]<=buf[n])
  @             && (n+1<=k<=bufLength-1 ==> buf[k]>=buf[n]))
  @    && \result==buf[n] ;
  @*/
public final static float findNthLowestNumber
               (float[] buf, int bufLength, int n) {
    int i,j;
    int l=0;
    int m=bufLength-1;
    float med=buf[n];
    float dum ;
    while (l<m) {
        i=l ;
        j=m ;
        do {
            while (buf[i]<med) i++ ;
            while (med<buf[j]) j-- ;
            dum=buf[j];
            buf[j]=buf[i];
            buf[i]=dum;
            i++ ; j-- ;
        } while ((j>=n) && (i<=n)) ;
        if (j<n) l=i ;
        if (n<i) m=j ;
        med=buf[n] ;
    }
    return med ;
}
\end{verbatim}

\normalsize

Given an array \texttt{\small buf} and two integers \texttt{\small bufLength} and \texttt{\small n},
the Fiji method \texttt{\small findNthLowestNumber} returns the $(n+1)$-th lowest number in the first
$\texttt{\small bufLength}$ components of \texttt{\small buf}. The precondition expresses that \texttt{\small buf}
is not null, \texttt{\small bufLength} must be an integer between $1$ and the length of \texttt{\small buf}, and \texttt{\small n}
is an integer between $0$ and $\texttt{\small bufLength}-1$. The definition of the postcondition includes the use of the predicate
\texttt{\small Permut}, a predefined predicate, which expresses that when the method returns the (modified) $\texttt{\small bufLength}$ first
components of the array \texttt{\small buf} must be
a permutation of the initial ones. The array has been reordered such that the components $0,\ldots,n-1$ are smaller than
or equal to the component $n$, and the elements at positions $n+1,\ldots, \texttt{\small bufLength}-1$ are greater than or equal to that in $n$.
The returned value must be equal to \texttt{\small buf[n]}, which is therefore the $(n+1)$-th lowest number in  the first
\texttt{\small bufLength} components of \texttt{\small buf}.

In order to prove the correctness of this program, we have included different JML annotations in the Java code.
First of all, loop invariants must be given for all \texttt{\small while} and \texttt{\small do} structures appearing in the code.
Difficulties have been found in order to deduce the adequate properties for invariants which
must be strong enough to imply the program (and other loops) postconditions; automated techniques like discovery of loop invariants~\cite{Ireland97onthe} will be 
used in the future. We show as an example
the loop invariant (and variant) for the exterior while, which is given by the following properties:

\footnotesize
\begin{verbatim}
/*@ loop_invariant
  @ 0<=l<=n+1 && n-1<=m<=bufLength-1 && l<=m+2
  @ && (\forall integer k1 k2; (0<=k1<=n && m+1<=k2<=bufLength-1)
  @         ==> buf[k1]<=buf[k2])
  @ && (\forall integer k1 k2; (0<=k1<=l-1 && n<=k2<=bufLength-1)
  @         ==> buf[k1]<=buf[k2])
  @ && Permut{Pre,Here}(buf,0,buf.length-1) && med==buf[n]
  @ && ((l<m)==> ((l<=n)&&(m>=n)));
  @ loop_variant m - l+2;
  @*/
\end{verbatim}
\normalsize

To help the automated provers to verify the program and prove the generated proof obligations it is also necessary
to introduce several assertions in some intermediate points of
the program and to use ghost variables which allow the system to deduce that the loop variant decreases.

Our final specification of this method 
includes 78 lines of JML annotations (for only 24 Java code lines).
Krakatoa/Why produces 175 proof obligations expressing the validity of the program. The automated theorem prover
Alt-Ergo is able to demonstrate all of them, although in some cases more than a minute (in an ordinary computer)
is needed; another prover included in Krakatoa, CVC3, is, on the contrary, only capable of proving 171.
The proofs of the lemmas
obtained by means of Alt-Ergo certify the correctness of the method with respect to the given specification.







In this particular example, the automated theorem provers integrated in Krakatoa are enough
to discharge all the proof obligations. In other cases, some properties are not proven, and
then one should try to prove them using \emph{interactive} theorem provers, as Coq. In this
architecture, we also introduce the ACL2 theorem prover, as explained in the next section.


\section{The role of ACL2}\label{sec:role}

In this section, we present the role played by ACL2 in our infrastructure to verify the
correctness of Java programs. The Why platform relies on automated provers, such as Alt-Ergo
or CVC3, and interactive provers, such as Coq or PVS, to discharge proof obligations;
however, it does not consider the ACL2 theorem prover to that aim. We believe that the use of
ACL2 can help in the proof verification process. The reason is twofold.

\begin{itemize}
 \item The scope of automated provers is smaller than the one of ACL2; therefore, ACL2
       can prove some of the proof obligations which cannot be discharged by automated provers.
 \item Moreover, interactive provers lack automation; then, ACL2 can automatically discharge
       proof obligations which would require user interaction in interactive provers.
\end{itemize}

We have developed \emph{Coq2ACL2}, a Proof General extension, which integrates ACL2 in our
infrastructure to verify Java programs; in particular, we work with ACL2(r) a variant of ACL2
which supports the real numbers~\cite{ACL2r} -- the formalisation of real analysis in theorem provers 
is an outstanding topic, see~\cite{Reals}. Coq2ACL2 features three main functions:

\begin{itemize}
 \item[] \textbf{F1.} it transforms Coq statements generated by Why to ACL2;
 \item[] \textbf{F2.} it automatically sends the ACL2 statements to ACL2; and
 \item[] \textbf{F3.} it displays the proof attempt generated by ACL2.
\end{itemize}

\noindent If all the statements are proved in ACL2; then, the verification process is ended.
Otherwise, the statements must be manually proved either in Coq or ACL2.

The major challenge in the development of Coq2ACL2 was the transformation of Coq statements to ACL2.
There is a considerable number of proposals documented in the literature related to the area of theorem
proving interoperability. We have not enough space here to do a thorough review, but we can classify the 
translations between proof assistants in two groups: \emph{deep}~\cite{GKR11,JBDD11,C12} and \emph{shallow}~\cite{KW11,OS06,D00}.

In our work, we took advantage of a previous shallow development presented in~\cite{XLL}, where
a framework called \emph{I2EA} to import Isabelle/HOL theories into ACL2 was introduced. The approach followed in~\cite{XLL}
can be summarized as follows. Due to the different nature of Isabelle/HOL and ACL2, it is not feasible to
replay proofs that have been recorded in Isabelle/HOL within ACL2. Nevertheless, Isabelle/HOL
statements dealing with first order expressions can be transformed to ACL2; and then, they can be used as a schema to guide the proof in ACL2.

\begin{figure}[h]
\centering
\begin{tikzpicture}
\draw[rounded corners] (-1,1) rectangle (1,2);
\draw (0,1.5) node{Isabelle/HOL};

\draw[rounded corners] (-1,-2) rectangle (1,-1);
\draw (0,-1.5) node{Coq};

\draw[rounded corners] (3,-.5) rectangle (5,.5);
\draw (4,0) node{XLL};

\draw[rounded corners] (7,-.5) rectangle (9,.5);
\draw (8,0) node{ACL2};

\draw[-latex] (1,-1.5) -- (3,-0.2);
\draw[-latex] (1,1.5) -- (3,0.2);
\draw[-latex] (5,0) -- (7,0);
\end{tikzpicture}
\caption{(Reduced) Architecture of the I2EA framework integrating Coq.}
\label{f:architecture}
\end{figure}
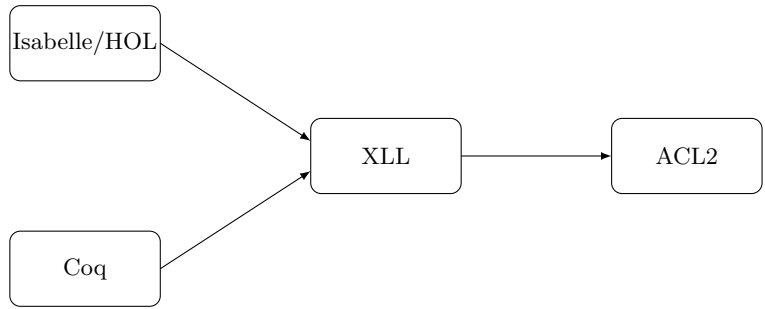

A key component in the framework presented in~\cite{XLL} was an XML-based specification language
called \emph{XLL} (that stands for Xmall Logical Language). XLL was developed to act as an intermediate
language to port Isabelle/HOL theories to both ACL2 and an Ecore model (given by UML class definitions and OCL
restrictions) -- the translation to Ecore serves as a general purpose formal specification of the theory
carried out. The transformations among the different languages are done by means of XSLT and some Java
programs. We have integrated the Coq system into the I2EA framework as can be seen in Figure~\ref{f:architecture}; in this way,
we can reuse both the XLL language and some of the XSLT files developed in~\cite{XLL} to transform (first-order like) Coq statements to ACL2.

In particular, functionality \textbf{F1} of Coq2ACL2 can be split into two steps:

\begin{enumerate}
 \item given a Coq statement, Coq2ACL2 transforms it to an XLL file using a Common Lisp translator program; then,
 \item the XLL file is transformed to ACL2 using an XSLT file previously developed in~\cite{XLL}.
\end{enumerate}

In this way, ACL2 has been integrated into our environment to verify Java programs. As we will see in the following
section, this has meant an improvement to automatically discharge proof obligations.


%


\section{The method in action: a complete example}\label{sec:action}

In our work, we deal with images acquired by microscopy techniques from
biological samples. These samples have volume and the object of interest is not always in the same plane.
For this reason, it is necessary to obtain  different planes from the same sample to get more information.
This means that several images are acquired in the same $XY$ plane at different levels of $Z$.
To work with this stack of images, it is often necessary to make their \emph{maximum projection}.
To this aim, Fiji provides several methods such as
maximum intensity or standard deviation to obtain the maximum projection of a set of images.

In this section, we consider the Fiji code for computing the maximum projection of a set of images based on
the standard deviation, which uses in particular the method \texttt{\small calculateStdDev} located
in the class \emph{ImageStatistics}.

\footnotesize
 \begin{verbatim}
double calculateStdDev(double n, double sum, double sum2) {
    double stdDev = 0.0;
    if (n>0.0) {
       stdDev = (n*sum2-sum*sum)/n;
       if (stdDev>0.0)
          stdDev = Math.sqrt(stdDev/(n-1.0));
       else
          stdDev = 0.0;
    } else
    stdDev = 0.0;
}
 \end{verbatim}

\normalsize
The inputs are \texttt{\small n} (the number of data to be considered), \texttt{\small sum}
(the sum of all considered values; in our case, these values will obtained from the pixels in an image) and
\texttt{\small sum2} (the sum of the squares of the data values). The method \texttt{\small calculateStdDev}
computes the standard deviation from these inputs and assigns it to the field \texttt{\small stdDev}.
The specification of this method is given by the following JML annotation.

\footnotesize
\begin{verbatim}
/*@ requires ((n==1.0)==> sum2==sum*sum) && ((n<=0.0) || (n>=1.0)) ;
  @ behaviour negative_n :
  @	  assumes  n<=0.0 || (n>0.0 && (n*sum2-sum*sum)/n <=0.0);
  @	  ensures stdDev == 0.0;
  @ behaviour normal_behaviour :
  @	  assumes n>=1.0 && ((n*sum2-sum*sum)/n  > 0.0);
  @	  ensures is_sqrt(stdDev,(double)((n*sum2-sum*sum)/n/(n-1.0)));
@*/
\end{verbatim}

\normalsize
The precondition, introduced by the keyword \texttt{\small requires}, expresses that in the case $\texttt{\small n}=1$ (that is, there is only one element in the data)
the inputs \texttt{\small sum} and \texttt{\small sum2} must satisfy $\texttt{\small sum2}=\texttt{\small sum}*\texttt{\small sum}$. Moreover we must require that 
\texttt{\small n} is less than or equal to $0$ or greater than or equal to $1$ to avoid the possible values in the interval $(0,1)$; for \texttt{\small n} in this interval
one has $\texttt{\small n}-1<0$ and then it is not possible to apply the square root function to the given argument $\texttt{\small stdDev}/(\texttt{n}-1.0)$. This fact has 
not been taken into account by the author of the Fiji program because in all real applications the method will be called with \texttt{\small n} being a natural number; however,
to formalise the method we must specify this particular situation in the precondition. For the postcondition we distinguish two different behaviours: if \texttt{\small n} is
non-positive or \texttt{\small sum} and \texttt{\small sum2} are such that  $\texttt{\small n}*\texttt{\small sum2}-\texttt{\small sum}*\texttt{\small sum}<0$, the field
\texttt{\small stdDev} is assigned to $0$; otherwise, the standard deviation formula is applied and the result is assigned to the field \texttt{\small stdDev}. The predicate
\texttt{\small is\_sqrt} is previously defined.

For the proof of  correctness of the method \texttt{\small calculateStdDev} in Krakatoa, it is necessary to specify (and verify) the method \texttt{\small sqrt}.
The problem here, as already explained in Section~\ref{sec:fijikrakatoa}, is that the method \texttt{\small sqrt} of the class \emph{Math} simply calls the equivalent method in the class
\emph{StrictMath}, and the code in \emph{StrictMath} of the method \texttt{\small sqrt} is just a native call and might be implemented differently on different Java platforms.
In order to give a JML specification of the method \texttt{\small sqrt} is necessary then to rewrite it with our own code. The documentation of \emph{StrictMath}
states ``\emph{To help ensure portability of Java programs, the definitions of some of the numeric functions in this package require that they produce the same results
as certain published algorithms. These algorithms are available from the well-known network library netlib as the package ``Freely Distributable Math Library'', fdlibm}''.
In the case of the square root, one of these \emph{recommended} algorithms is Newton's method; based on it, we have implemented and specified in JML
the computation of the square root of a given (non-negative) input of type double.

\footnotesize
\begin{verbatim}
/*@ requires c>=0 && epsi > 0 ;
  @ ensures \result >=0 &&  (\result*\result>=c)
  @	   &&	\result*\result - c < epsi ;
  @*/
public double sqrt(double c, double epsi){
    double t;
    if (c>1) t= c;
        else t=1.1;
    /*@ loop_invariant
      @ (t >= 0) && (t*t> c) ;
      @*/
    while (t* t - c  >= epsi) {
        t = (c/t + t) / 2.0;
    }
    return t;
}
	
/*@ requires c>=0  ;
  @ ensures (\result >=0) && (\result*\result>=c)
  @ && (\result*\result - c < 1.2E-7);
  @*/
public double sqrt(double c){
    double eps=1.2E-7;
    return sqrt(c,eps);
}
\end{verbatim}
\normalsize

The first method computes the square root of a double \texttt{\small x} with a given precision \texttt{\small epsi}; the second one
calls the previous method with a precision less than $1.2E-7$.
Using JUnit, we have run one million tests between $1E9$ and $1E-9$ to show that the results of our method \texttt{\small sqrt}
have similar precision to those obtained by the \emph{original} method \texttt{\small Math.sqrt}. Here, we applied the
``first test, then verify'' approach -- intensive testing can be really useful to find bugs (and can save us time) before starting the
verification process.

From the given JML specification for the Fiji method \texttt{\small calculateStdDev} and our \texttt{\small sqrt} method, Why/Krakatoa
produces 52 proof obligations, 9 of them corresponding to lemmas that we have introduced and which are used in order to prove the correctness of the programs.
Alt-Ergo is able to prove 50 of these proof obligations, but two of the lemmas that we have defined remain unsolved. CVC3 on the contrary only proves 44 proof obligations.

The two lemmas that Alt-Ergo (and CVC3) are not able to prove are the following ones:

\footnotesize
\begin{verbatim}
/*@ lemma double_div_pos :
  @ \forall double x y; x>0 && y > 0 ==> x / y > 0;
  @*/
/*@ lemma double_div_zero :
  @ \forall double x y; x==0.0 && y > 0 ==> x / y == 0.0;
  @*/
\end{verbatim}
\normalsize

In order to discharge these two proof obligations, we can manually prove their associated Coq expressions.

\footnotesize
\begin{verbatim}
Lemma double_div_zero : (forall (x_0_0:R), (forall (y_0:R),
  ((eq x_0_0 (0)%R) /\ (Rgt y_0 (0)%R) -> (eq (Rdiv x_0_0 y_0) (0)%R)))).

Lemma double_div_pos : (forall (x_13:R), (forall (y:R),
  ((Rgt x_13 (0)%R) /\ (Rgt y (0)%R) -> (Rgt (Rdiv x_13 y) (0)%R)))).
\end{verbatim}
\normalsize

\begin{figure}
 \centering
 \includegraphics[scale=.35]{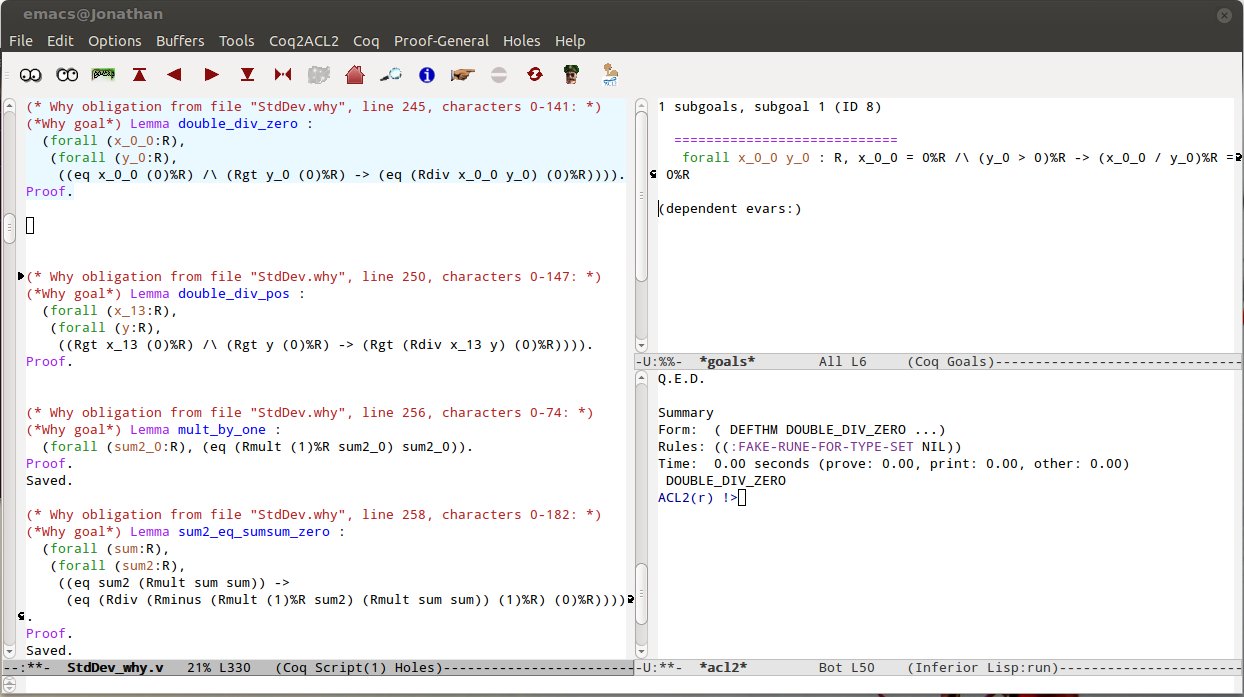}
 \caption{Proof General with Coq2ACL2 extension. The Coq2ACL2 extension consists of the Coq2ACL2 menu and the right-most button of the toolbar.
 Left: the Coq file generated by the Why tool. Top Right: current state of the Coq proof. Bottom Right: ACL2 proof of the lemma.}\label{fig:coq2acl2}
\end{figure}

Both lemmas can be proven in Coq in less than 4 lines, but, of course, it is necessary some experience working
with Coq. Therefore, it makes sense to delegate those proofs to ACL2. Coq2ACL2 translates the Coq lemmas to the
following ACL2 ones. ACL2 can prove both lemmas without any user interaction (a screenshot of the proof of one of this
lemmas in ACL2 is shown in Figure~\ref{fig:coq2acl2}).

\footnotesize
\begin{verbatim}
(defthm double_div_zero
 (implies (and (realp x_0_0) (realp y_0) (and (equal x_0_0 0) (> y_0 0)))
          (equal (/ x_0_0 y_0) 0)))

(defthm double_div_pos
 (implies (and (realp x_13) (realp y) (and (> x_13 0) (> y 0)))
          (> (/ x_13 y) 0))
\end{verbatim}
\normalsize

\section{Conclusions and further work}\label{sec:conclusions}

This paper reports an experience to verify actual Java code,
as generated by different-skilled programmers, in a multi-programmer tool
called Fiji. As one could suspect, the task is challenging and,
in some sense, the objectives are impossible to accomplish, at least
in their full extent -- after our experiments, we have found
that the Fiji system is \emph{unsound}, but the errors are minor (e.g. a variable declared as a real number but which should be declared as an
integer) and can be easily corrected.

Nevertheless, we defend the interest of this kind of experimental work. It
is useful to evaluate the degree of maturity of the verification tools (Krakatoa,
in our case). In addition, by a careful examination of the code really needed for a concrete
application, it is possible to isolate the relevant parts of the code, and then it is
possible to achieve a complete formalisation. Several examples in our text showed this
feature, see Section~\ref{sec:spec}.

In addition to Krakatoa, several theorem provers (Coq and ACL2) have been used
to discharge some proof obligations that were not automatically proved by Krakatoa. To this aim, 
it has been necessary the integration of several tools, and our approach can 
be considered as semi-formal: we keep transformations as simple as possible, and substantiate the process 
by systematic testing.

As a further interest of our work, we have reused a previous interoperability-proposal~\cite{XLL},
between Isabelle and ACL2, to get an integration of ACL2 (through a partial mapping from Coq to ACL2),
without touching the Krakatoa kernel.

Future work includes several improvements in our method. Starting from the beginning, the transformation
from real Java code to Krakatoa one could be automated (Section~\ref{sec:fijikrakatoa} can be understood as a list of requirements
to this aim). Then, a formal study of this transformation could be undertaken to increase the reliability
of our method. In addition, we can try to automatically reconstruct ACL2 proofs in Coq.

As for applications, more verification is needed to obtain a certified version of, for instance, the
SynapCountJ plug-in~\cite{synapcount}. The preliminary results presented in this paper allow us to
be reasonably optimistic with respect to the feasibility of this objective.


\bibliographystyle{plain}
\bibliography{vpdims}

\end{document}